\documentclass[letter,nofootinbib,12pt]{revtex4-2}

\usepackage[colorlinks,citecolor=blue]{hyperref}
\usepackage{amsmath}
\usepackage{mathrsfs}
\usepackage{bbm}
\usepackage{amsfonts}
\usepackage{amssymb}
\usepackage{latexsym}
\usepackage{graphicx}
\usepackage[english]{babel}
\usepackage{multirow}
\usepackage{float}
\usepackage{url}
\usepackage{slashed}
\usepackage[compat=1.1.0]{tikz-feynman}
\usepackage{orcidlink}

\renewcommand{\Delta}{\varDelta} 
\renewcommand{\Gamma}{\varGamma} 
\renewcommand{\Omega}{\varOmega} 
\renewcommand{\Phi}{\varPhi} 
\renewcommand{\Psi}{\varPsi} 
\renewcommand{\Sigma}{\varSigma} 
\renewcommand{\Theta}{\varTheta} 
\renewcommand{\epsilon}{\varepsilon}

\newcommand{\be}{\begin{equation}}
\newcommand{\ee}{\end{equation}}
\newcommand{\ba}{\begin{array}}
\newcommand{\ea}{\end{array}}
\newcommand{\bea}{\begin{eqnarray}}
\newcommand{\eea}{\end{eqnarray}}

\begin{document}

\title{The Amaterasu Cosmic Ray as a Magnetic Monopole and Implications for Extensions of the Standard Model}

\author{Paul H. Frampton}
\email{paul.h.frampton@gmail.com}
\affiliation{Dipartimento di Matematica e Fisica “Ennio De Giorgi”,
Universit`a del Salento and INFN-Lecce,
Via Arnesano, 73100 Lecce, Italy} 
\author{Thomas W. Kephart \orcidlink{0000-0001-6414-9590} \, }
\email{tom.kephart@gmail.com}
\affiliation{Department of Physics and Astronomy, Vanderbilt
University, Nashville, TN 37235, USA} 
 

\begin{abstract}
\noindent
The Amaterasu cosmic ray particle appears to have come from the direction of the local cosmic void.
We take this as evidence that it is a magnetic monopole rather than a proton or nucleus. This in turn 
strongly suggests physics at high energy  is described by a quiver gauge theory.  
\end{abstract}


\maketitle

\newpage

\noindent
{\it Introduction}:

The Greisen–Zatsepin–Kuzmin \cite{Greisen:1966jv,Zatsepin:1966jv} limit $\sim 5\times 10^{19}$ eV ( or GZK cutoff) is the upper bound on the energy of cosmic ray protons traveling from their production in distant galaxies through the intergalactic medium to us. The limit is due to cosmic ray proton energy being degraded by interacting with cosmic microwave background photons via the process $p+\gamma\rightarrow \Delta^* \rightarrow p+\pi^0$. The mean free path (mfp) for this process is  $\sim 6$ Mpc. The recent Amaterasu cosmic ray particle found by the Telescope Array Group \cite{TelescopeArray:2023sbd} appears to have come from the direction of the local cosmic void, which subtends a solid angle of approximately $1.6\, \pi$ and is at least 45 Mpc across  \cite{TullyFisher,Tully:2019ngb}, so it is unlikely that a cosmic ray proton could cross such a distance unscathed. (To make it all the way across the local void the particle would need an initial energy of about $10^{23}$ eV which seems highly unlikely, given known acceleration mechanisms.) That leads us to conclude the following, either (i) there is a source of ultra high energy cosmic rays (UHECRs) within the void, (ii) the path of the  Amaterasu particle was bent substantially by a foreground object, presumably within the Milky Way (see e.g., \cite{Anchordoqui:2018qom}), or (iii) the Amaterasu particle is not a proton or nucleus, but instead a magnetic monopole (MM) \cite{Kephart:1995bi,Wick:2000yc}. We can think of free MMs as having no galactic source, but rather as particles that have wondered the Universe since their production in an early gauge theory phase transition.

Option (i), a UHECR source is possible, but appears unlikely and has been analysed in detail by \cite{Unger:2023hnu}. While there are a few dwarf galaxies in the local void, the chances of producing UHECRs there is diminished by the ratio
of the density of galaxies in the local group compared to the density off galaxies in the local void, roughly a factor of $10^3$, which renders option (i) unlikely. Option (ii), could be a UHECR proton or nucleus that appears to have originated in the local void, but in actuality was produced elsewhere, say in the local group, and who's path was bent by interacting with a local object containing a strong 
extended B-field to make it appear as if the particle came from the local void. Again we consider this possibility unlikely. Details of an analysis along these lines can be found in the extensive review by Anchordoqui \cite{Anchordoqui:2018qom}. For other constraints on nearby sources of ultra high energy cosmic rays see \cite{Kuznetsov:2023jfw}.

 The probability of bending such a high energy particle in the foreground, i.e., option (ii) would require an encounter with a strong magnetic field over a length scale L. Details can be extracted for such estimates from  \cite{Anchordoqui:2018qom}. Again this is highly unlikely and leads us to conclude that option (iii), that the Amaterasu particle was a magnetic monopole,  is our best choice.

If we assume a magnetic monopole  is the solution to the mystery of the  Amaterasu particle, then we need to discuss the 
necessary properties of such a monopole and in what theories it might arise. First the monopole must be relativistic, which means it is relatively light. For example, if the  monopole arose in an $SU(5)$ grand unified theory, either SUSY or nonSUSY, then its
mass would be $M_{GUT}/\alpha$ or $\sim 10^{17}$ GeV, or $\sim 10^{26}$ eV, far too heavy to be relativistic. If we require a relativistic
gamma factor of say, at least $10^3$, then the monopole mass should be roughly $10^{17}$ eV  ($10^5$ TeV) or less. That puts a strong constraint on models where magnetic monopoles appear. However, there is a large class of such models that we will discuss.

Requiring that monopoles not consume
galactic magnetic fields faster than galactic dynamos can
regenerate them results in the Parker bound is
of $10^{-15}$/$\rm{cm}^2$/s/sr. The Parker bound is several orders of magnitude
above the observed highest--energy cosmic ray flux of 1 per km$^2$ per century, 
hence does not constrain UHECRs.
\vspace{0.5cm}

\noindent
{\it Monopole Energy:}

The kinetic energy gained by magnetic monopole on traveling along a magnetic
field of coherence length $\xi$ is \cite{Kephart:1995bi}
\begin{equation}
E
\sim\,g\,B\,\xi\, 
\label{Ekin}
\end{equation}
where
\begin{equation}
g=e/2\alpha=3.3\times10^{-8} \;\rm{esu} \;\;
({\rm or}\; 3.3\times 10^{-8} dynes/G)
\label{charge}
\end{equation}
is the magnetic charge according to the Dirac quantization condition,
$B$ is the magnetic field strength, $\xi$ is the
field's coherence.
Allowing the monopole to random--walk through the $n$ domains
of coherent fields would increase the result by roughly $\sqrt{n}$.
Typical astrophysical magnetic fields in galaxies, galaxy cluster, AGN jets, etc. range from 0.1--100 $\mu G$ while the coherence length of these fields range from $10^{-4}-30$ Mpc. This leads to monopole energies in the range 
$$1.7\times 10^{20}\,\, to\,\, 5\times 10^{23}\,\, eV.$$
See \cite{Wick:2000yc} for more details, and for a more recent comprehensive study of the acceleration of monopoles in intergalactic magnetic fields see \cite{Perri:2023ncd}.

It is important to note that the current bound on the flux of relativistic monopoles from Ice Cube is \cite{IceCube:2021eye}
$$F_{_{MM}}\le 2\times 10^{-19} cm^{-2}s^{-1}sr^{-1},$$
while the observed flux of UHECRs is only about one per $km^2$ per century, or
$$F_{_{UHECR}} \sim 3\times 10^{-20} cm^{-2}s^{-1}sr^{-1}.$$

The Amaterasu event had energy $E_A = 2.44 \times 10^{20}$ eV and so falls within the expected monopole energy range, and it is also not disfavored by the Ice Cube bound. 

The highest initial energy UHECR ever seen is the Fly's Eye event \cite{HIRES:1994ijd} at energy $E_{FE} = 3.2 \times 10^{20}$ eV. The Amaterasu event and two events from AGASA \cite{AGASA,Shinozaki:2006kk} at
nominal energies $E_{A1} = 2.46 \times 10^{20}$ eV and
$E_{A2} = 2.13 \times 10^{20}$ eV are all within 1 $\sigma$ of each other, hence within errors of being second in energy to
the Fly's Eye  event.  AGASA reported eleven events in all with energy above $10^{20}$ eV. 
The Pierre Auger experiment has reported there 20 highest energy  events range between $1.10$ and $1.66 \times 10^{20}$ eV \cite{PierreAuger:2022qcg,PierreAuger:2022axr}.

\vspace{0.5cm}
\noindent
{\it Monopole Direction:}

The direction of the center of the local void is \cite{TullyFisher,Tully:2019ngb}

(RA, Dec) =  $(279.5^{\circ},18.0^{\circ},)$

\noindent
while the arrival direction of the Amaterasu comic ray was 

(RA, Dec) = $(255.9^{\circ}, 16.1^{\circ})$

\noindent
which is well within the direction of the local void. Even if backtracked through the galactic field, a proton or nucleus still
appears to originate within the local void \cite{Unger:2023hnu}.
Given that there are no apparent sources of UHECR protons or nuclei within the local void and that it is highly unlikely that such a trajectory could have been bent by a foreground object within our galaxy, we are led to conclude that the Amaterasu particle was a magnetic monopole. Since it was relativistic this constrains the monopole mass 
$$
M \le 10^{8} \,\, GeV.
$$
We can also arrive at the following lower bound on the monopole mass by requiring the phase transition where monopoles are produced to be above the electroweak scale
$$
M \ge 10^{4} \,\, GeV.
$$
This makes it difficult for the symmetry breaking where monopoles were produced to be of standard grand unification type, i.e, in the $SU(5), \, SO(10), E_6$ chain and more likely to be a quiver gauge theory of the type which we now discuss.

\vspace{0.5cm}
\noindent
{\it Models with Light Magnetic Monopoles:} 

The general class of quiver gauge theories are most easily arranged to have light magnetic monopoles since they can  have separated gauge and flavor groups, and hence avoid proton decay with a low symmetry breaking scale where the $U(1)$ appears that is associated with the monopole. The simplest examples of this are the Pati-Salam model (PS) \cite{Pati:1974yy} and the Trinification model (T) \cite{trin} with gauge groups $SU(4)\times SU(2)\times SU(2)$ and $SU(3)\times SU(3)\times SU(3)$ respectively.

The appearance of monopoles in quiver gauge theories
which generalise the original PS and T models has
been analysed in \cite{Kephart:2001ix,Kephart:2006zd,Kephart:2017esj,Sheridan:2022qku}. The last two of these papers used the
LieART platform \cite{Feger:2012bs,Feger:2019tvk}
to study, as the simplest extensions of PS and T,
gauge groups with three factors $G =SU(a) \times SU(b) \times SU(c)$ with $a \geq 3, b, c \geq 2$ subject to
an upper limit $\dim[G]= (a^2 + b^2 + c^2 -3) \leq 78 =\dim[E(6)]$.
By restricting the size of $G$, a manageably finite, but large, number
of models are selected and studied more completely analysed than was previously accomplished by hand. 
Further papers on quiver gauge theories include
\cite{Kim:1980yk,Kachru:1998ys,Lawrence:1998ja,Frampton:2000zy,Frampton:2000mq,Frampton:2007fr}.
 
Initially, the chiral fermions are all put into bifundamental representations,
then permissible additional anomaly-free non-bifundamental combinations
are added. For example, in an $(abc) = (333)$ theory, one can add
$3(3,1,1)+3(1,{\bar 3},1)+({\bar 3},3,1)$ while keeping anomaly cancellation,
although the theory is no longer a subgroup of an $E_6$ grand unified theory (GUT).

Once we depart from the framework of GUTS, leptons with fractional
electric charges {\it e.g.} $(\pm \frac{e}{2}, \pm \frac{e}{6})$,
appear. Thus, if the Amaterasu
cosmic ray is a monopole, it suggest the probable existence of fractionally-charged
leptons in particle theory \cite{Kephart:2001ix,Kephart:2006zd,Kephart:2017esj,Sheridan:2022qku}. Fractional electric charged particles
in turn predicts multiplely charged monopoles in order that the Dirac
relation $ge=n\hbar/2$ is maintained.

The mass of the magnetic monopoles can naturally be at an intermediate
scale such as the expected mass of the Amaterasu particle and new particle physics, 
like fractional charged lepton,  should be expected
at a few $TeV$ scale. Indirect evidence of monopoles may be accessible to the LHC 
where there are now dedicated monopole searches such as the MoEDAL 
experiment \cite{MoEDAL:2009jwa,MoEDAL:2014ttp}.
Searches for the additional light particles, such as fractionally-charged
leptons as predicted by such magnetic monopole theories, can be performed
at the Large Hadron Collider (LHC). 

Since we want a high energy monopole to look like the Amaterasu event, our interest is in baryonic monopoles, bound states of monopoles with both color magnetic and $U(1)_{EM}$ magnetic charge. These particles exist in various quiver gauge theories, but for
the present discussion we focus on a model independent analysis. For a hard collision involving one of the constituent quarks in normal baryon, the quark winds up on the end of a QCD electric flux tube that typically breaks once it is long enough to fragment into mesons.
If the initial collision is between a proton and a nucleus in our upper atmosphere, then a cascade follows and the remnants can be seen in a detector on the surface of the earth.
For a collision involving an individual monopole within the baryonic monopoles, a string (color magnetic flux tube) forms, but it can not break unless there is enough energy
to produce a monopole-antimonople pair. Consequently in most such collisions the color magnetic string oscillates and radiates light particles. But while this string is still stretched, the 
excited baryonic monopole's cross section is dramatically increased and further scattering
takes place that can mimic the air shower of a ultra high energy proton or nucleus \cite{Wick:2000yc}.

\vspace{0.5cm}
\noindent
{\it Conclusion}: 

Cosmic rays have historically played a major r\^{o}le in particle physics, such as the original discoveries of the positron and the muon. The Amaterasu cosmic ray is only one event but
it is an extraordinary one. It is one of the most energetic
cosmic rays ever recorded and the only one of those pointing
back to the local void where there is no obvious source.

This renders it unlikely that the primary is a proton or nucleus
and leads to our favoured interpretation as the long-sought
magnetic monopole predicted in beautiful theory invented by
Dirac in 1931. (That the Amaterasu particle may be a magnetic monopole has also recently been suggested by \cite{Cho:2023krz}.) If this is correct, it impacts on what is the
most likely extension of the standard model of particle theory,
for which we have suggested a quiver gauge theory with chiral fermions in bifundamental representations. This also predicts the probable existence of fractionally
charged leptons which could be discovered at the LHC.
The remarkable and unexpected observation of the
Amaterasu cosmic ray, reported first in November 2023, provides a potentially
revolutionary step forward in the theory of particle physics.

 \noindent
{\it Acknowledgement}:

We dedicate this work to the memory of our friend and colleague Tom Weiler, who had an abiding interest in cosmic rays and magnetic monopoles.


\begin{thebibliography}{999}

\bibitem{Greisen:1966jv}
  K.~Greisen,
  Phys.\ Rev.\ Lett.\  {\bf 16}, 748 (1966)
  doi:10.1103/PhysRevLett.16.748.

\bibitem{Zatsepin:1966jv}
  G.~T.~Zatsepin and V.~A.~Kuzmin,
  JETP Lett.\  {\bf 4}, 78 (1966)
  [Pisma Zh.\ Eksp.\ Teor.\ Fiz.\  {\bf 4}, 114 (1966)].

\bibitem{TelescopeArray:2023sbd}
R.~U.~Abbasi \textit{et al.} [Telescope Array],
Science \textbf{382}, 903-907 (2023)
doi:10.1126/science.abo5095
[arXiv:2311.14231 [astro-ph.HE]].

\bibitem{TullyFisher}
R. B. Tully and J.R. Fisher, ``Nearby Galaxies Atlas,''
Cambridge University Press, 1987.

\bibitem{Tully:2019ngb}
R.~B.~Tully, D.~Pomarede, R.~Graziani, H.~M.~Courtois, Y.~Hoffman and E.~J.~Shaya,
Astrophys. J. \textbf{880}, no.1, 24 (2019)
doi:10.3847/1538-4357/ab2597
[arXiv:1905.08329 [astro-ph.CO]].



\bibitem{Anchordoqui:2018qom}
L.~A.~Anchordoqui,
Phys. Rept. \textbf{801}, 1-93 (2019)
doi:10.1016/j.physrep.2019.01.002
[arXiv:1807.09645 [astro-ph.HE]].

\bibitem{Kephart:1995bi}
T.~W.~Kephart and T.~J.~Weiler,
Astropart. Phys. \textbf{4}, 271-279 (1996)
doi:10.1016/0927-6505(95)00043-7
[arXiv:astro-ph/9505134 [astro-ph]].

\bibitem{Wick:2000yc}
S.~D.~Wick, T.~W.~Kephart, T.~J.~Weiler and P.~L.~Biermann,
Astropart. Phys. \textbf{18}, 663-687 (2003)
doi:10.1016/S0927-6505(02)00200-1
[arXiv:astro-ph/0001233 [astro-ph]].


\bibitem{Unger:2023hnu}
M.~Unger and G.~R.~Farrar,
Astrophys. J. Lett. \textbf{962}, no.1, L5 (2024)
doi:10.3847/2041-8213/ad1ced
[arXiv:2312.13273 [astro-ph.HE]].

\bibitem{Kuznetsov:2023jfw}
M.~Y.~Kuznetsov,
[arXiv:2311.14628 [astro-ph.HE]].



\bibitem{Perri:2023ncd}
D.~Perri, K.~Bondarenko, M.~Doro and T.~Kobayashi,
[arXiv:2401.00560 [hep-ph]].

\bibitem{IceCube:2021eye}
R.~Abbasi \textit{et al.} [IceCube],
Phys. Rev. Lett. \textbf{128}, no.5, 051101 (2022)
doi:10.1103/PhysRevLett.128.051101
[arXiv:2109.13719 [astro-ph.HE]].

\bibitem{HIRES:1994ijd}
D.~J.~Bird \textit{et al.} [HIRES],
Astrophys. J. \textbf{441}, 144-150 (1995)
doi:10.1086/175344
[arXiv:astro-ph/9410067 [astro-ph]].


\bibitem{AGASA}
See the AGASA website:
http://www-akeno.icrr.u-tokyo.ac.jp/AGASA/results.html\#highest

\bibitem{Shinozaki:2006kk}
K.~Shinozaki [AGASA],
Nucl. Phys. B Proc. Suppl. \textbf{151}, 3-10 (2006)
doi:10.1016/j.nuclphysbps.2005.07.002

\bibitem{PierreAuger:2022axr}
P.~Abreu, \textit{et al.} [Pierre Auger],
Astrophys. J. \textbf{935}, no.2, 170 (2022)
doi:10.3847/1538-4357/ac7d4e
[arXiv:2206.13492 [astro-ph.HE]].

\bibitem{PierreAuger:2022qcg}
A.~A.~ Halim, \textit{et al.} [Pierre Auger],
Astrophys. J. Suppl. \textbf{264}, no.2, 50 (2023)
doi:10.3847/1538-4365/aca537
[arXiv:2211.16020 [astro-ph.HE]].



\bibitem{Pati:1974yy}
J.~C.~Pati and A.~Salam,
Phys.\ Rev.\ D {\bf 10}, 275 (1974).

\bibitem{trin}
S. L. Glashow, Fifth Workshop on Grand Unification: Proceedings.  Edited
by Kyungsik Kang, Herbert Fried, Paul Frampton. World Scientific,
1984. 538p.


\bibitem{Kephart:2001ix}
T.~W.~Kephart and Q.~Shafi,
Phys.\ Lett.\ B {\bf 520}, 313 (2001)
[arXiv:hep-ph/0105237].

\bibitem{Kephart:2006zd}
T.~W.~Kephart, C.~A.~Lee and Q.~Shafi,
JHEP \textbf{01}, 088 (2007)
doi:10.1088/1126-6708/2007/01/088
[arXiv:hep-ph/0602055 [hep-ph]].

\bibitem{Kephart:2017esj}
T.~W.~Kephart, G.~K.~Leontaris and Q.~Shafi,
JHEP \textbf{10}, 176 (2017)
doi:10.1007/JHEP10(2017)176
[arXiv:1707.08067 [hep-ph]].

\bibitem{Sheridan:2022qku}
E.~Sheridan and T.~W.~Kephart,
Nucl. Phys. B \textbf{987}, 116108 (2023)
doi:10.1016/j.nuclphysb.2023.116108
[arXiv:2206.13309 [hep-ph]].

\bibitem{Feger:2012bs}
R.~Feger and T.~W.~Kephart,
Comput. Phys. Commun. \textbf{192}, 166-195 (2015)
doi:10.1016/j.cpc.2014.12.023
[arXiv:1206.6379 [math-ph]].

\bibitem{Feger:2019tvk}
R.~Feger, T.~W.~Kephart and R.~J.~Saskowski,
Comput. Phys. Commun. \textbf{257}, 107490 (2020)
doi:10.1016/j.cpc.2020.107490
[arXiv:1912.10969 [hep-th]].

\bibitem{Kim:1980yk}
J.~E.~Kim and H.~S.~Song,
Phys.\ Rev.\ D {\bf 22}, 1753 (1980).

\bibitem{Kachru:1998ys}
S.~Kachru and E.~Silverstein,
Phys.\ Rev.\ Lett.\ {\bf 80}, 4855(1998)
[hep-th/9802183].

\bibitem{Lawrence:1998ja}
A.~E.~Lawrence, N.~Nekrasov and C.~Vafa,
Nucl. Phys. B \textbf{533}, 199-209 (1998)
doi:10.1016/S0550-3213(98)00495-7
[arXiv:hep-th/9803015 [hep-th]].


\bibitem{Frampton:2000zy}
P.~H.~Frampton and T.~W.~Kephart,
Phys.\ Lett.\ B {\bf 485}, 403 (2000) [hep-th/9912028].

\bibitem{Frampton:2000mq}
  P.~H.~Frampton and T.~W.~Kephart,
  Phys.\ Rev.\ D {\bf 64}, 086007 (2001)
  [arXiv:hep-th/0011186].

\bibitem{Frampton:2007fr}
P.~H.~Frampton and T.~W.~Kephart,
Phys. Rept. \textbf{454}, 203-269 (2008)
doi:10.1016/j.physrep.2007.09.005
[arXiv:0706.4259 [hep-ph]].

\bibitem{MoEDAL:2009jwa}
J.~Pinfold \textit{et al.} [MoEDAL],
CERN-LHCC-2009-006.

\bibitem{MoEDAL:2014ttp}
B.~Acharya \textit{et al.} [MoEDAL],
Int. J. Mod. Phys. A \textbf{29}, 1430050 (2014)
doi:10.1142/S0217751X14300506
[arXiv:1405.7662 [hep-ph]].

\bibitem{Cho:2023krz}
Y.~M.~Cho and F.~H.~Cho,
[arXiv:2312.08115 [hep-ph]].


\end{thebibliography}
\end{document}